\documentclass[useAMS,usenatbib]{mn2e}
\usepackage{amssymb}
\usepackage{times}
\usepackage{epsfig}
\newif\ifAMStwofonts

\def\fig#1{Figure~\ref{#1}}
\def\Fig#1{Figure~\ref{#1}}

\def\tab#1{Table~\ref{#1}}
\def\sec#1{Section~\ref{#1}}

\newcommand{\luminsub}{$L_{850\mu m}$}

\newcommand{\subcolour}{$L_{850\mu m}/L_{2.2\mu m/(1+z)}$}

\newcommand{\aj}{AJ}
\newcommand{\apj}{ApJ}
\newcommand{\apjs}{ApJS}
\newcommand{\apjl}{ApJ}
\newcommand{\mnras}{MNRAS}

\newcommand{\nat}{Nature}
\newcommand{\aap}{A\&A}

\title[Coupled Spheroid and Black-Hole Formation]{Coupled spheroid and black-hole formation, and the multifrequency
detectability of active galactic nuclei and submillimetre sources.}

\author[E. N. Archibald et al.]{E. N. Archibald$^{1,2}$\thanks{email:
    e.archibald@jach.hawaii.edu}, J. S. Dunlop$^{2}$, R. Jimenez$^{3}$, A. C. S. Fria\c{c}a$^{4}$, R. J. McLure$^{2,5}$, \and and D. H. Hughes$^{6}$\\
$^1$Joint Astronomy Centre, 660 N. A`oh\={o}k\={u} Place, University Park, Hilo, HI, 96720 \\
$^2$Institute for Astronomy, University of Edinburgh, Royal Observatory, Edinburgh, EH9 3HJ, Scotland \\
$^3$Dept. of Physics \& Astronomy, Rutgers University, 136 Frelinghuysen Road, Piscataway, NJ, 08854-8091 \\
$^4$Instituto Astronomico e Geofisico, USP, Av. Miguel Stefano 4200, 04301-904 Sao Paulo, SP, Brazil \\
$^5$Astrophysics, Department of Physics, Keble Road, Oxford, OX1 3RH, England \\
$^6$Instituto Nacional de Astrofisica, Optica y Electronica (INAOE), Apartado Postal 51 y 216, 72000, Puebla, Pue., Mexico \\
}

\date{Accepted ;
      Received ;
      in original form }

\pagerange{\pageref{firstpage}--\pageref{lastpage}}
\pubyear{2001}

\begin{document} 
\label{firstpage}

%%%%% Make the title %%%%%

\maketitle 

%%%%% Abstract %%%%%

\begin{abstract} 
  
We use a simple model of spheroid formation to explore the relationship
between the creation of stars and dust in a massive proto galaxy and the 
growth of its central black hole. This model predicts that sub-mm luminosity 
peaks after only $\simeq 0.2$ Gyr. However, without a very massive seed black 
hole, Eddington-limited growth means that a black hole mass of $10^9 
\,{\rm M_{\odot}}$, and hence very luminous AGN activity, cannot be produced until 
$>0.5$ Gyr after the formation of the first massive stars in the
halo. The model thus predicts a time-lag between the peak of sub-mm
luminosity and AGN luminosity in a massive proto-elliptical of a few
times $10^8$ years. For a formation redshift $z \simeq 5$, this means
that powerful AGN activity is delayed until $z \simeq 3.5$, by which
time star formation in the host is $\simeq 90\%$ complete, and sub-mm
luminosity has declined to $\simeq 25\%$ of its peak value. This
provides a natural explanation for why successful sub-mm detections of
luminous radio galaxies are largely confined to $z > 2.5$.  Conversely
the model also predicts that while all high-redshift luminous
sub-mm-selected sources should contain an active (and growing) black
hole, the typical luminosity of the AGN in such objects is $\simeq
1000$ times smaller than that of the most powerful AGN. This is
consistent with the almost complete failure to detect sub-mm selected
galaxies with existing X-ray surveys. Finally the model yields a
black-hole:spheroid mass ratio which evolves rapidly in the first Gyr,
but asymptotes to $\simeq 0.001-0.003$ in agreement with results at
low redshift. This ratio arises not because the AGN terminates star
formation, but because fueling of the massive black hole is linked to
the total mass of gas available for star formation in the host.

\end{abstract}


%%%%% Introduction %%%%%

\section{Introduction}
\label{intro}
Until recently the formation/evolution of active galactic nuclei and the
formation/evolution of galaxies of stars have been viewed and studied as
separate problems in observational cosmology. However, two recent
observational discoveries have changed this. First, it is now clear that the
global history of star-formation and AGN activity track each other rather
well, at least out to $z \simeq 2$
\citep{dunlop1997,BoyleTerlevich1998,fran2000}. Second, it is also now well
established that essentially all present-day galaxies contain a central black
hole whose mass is proportional to the mass of the spheroidal component of the
host, with $M_{bh} \simeq 0.001 - 0.003\,M_{sph}$ \citep{kormendy2001,mf2001}.
Consequently the growth and fueling of super-massive black holes and the
construction of galactic spheroids are now seen as potentially
intimately-related processes
\citep[e.g.,][]{silkrees1998,fabian1999,granato2001}.

This increased awareness of possible links between black-hole and spheroid
formation has coincided with recent observational advances which have the
potential to provide greatly-improved constraints on the evolution of massive
galaxies and AGN. In particular new sub-mm surveys with the SCUBA camera on
the JCMT have resolved a substantial fraction of the sub-mm background
\citep{smailsurvey1997,hdfnature,ealessurvey1999}, and have revealed a
population of massive, dusty star-forming objects at $z > 1.5 - 2$ which are
sufficiently numerous to account for the formation of all present-day massive
ellipticals \citep{dunlop2001,susiesurvey}. At the same time, deep X-ray
surveys have resolved the X-ray background and have enabled the evolution of
X-ray emitting black holes to be traced out to $z \simeq 4$ \citep{mhs1999}.
Moreover, motivated by the potential links between black-hole and spheroid
formation outlined above, there is considerable current interest in
determining the level of overlap between the sub-mm and X-ray populations.
Current observations indicate this overlap is, perhaps surprisingly, small
$\simeq 10\%$ \citep{fabianxray2000,omarxray2002,bargerxray2001}.

An alternative route to exploring the link between black-hole and spheroid
formation is provided by observations designed to measure the star-formation
history of the hosts of known active black holes as a function of redshift.
\citet{rgmnpaper} performed this experiment for the most massive black holes
by undertaking a sub-mm study of a sample of $\simeq 50$ radio galaxies
spanning a wide range in redshift $1<z<5$ at constant radio power. At low
redshift it is now well established that powerful radio sources are produced
by only the most massive black holes with $M_{bh} > 10^9\,{\rm M_{\odot}}$, residing
in massive elliptical hosts with $M_{sph} \simeq 10^{12}\,{\rm M_{\odot}}$ \citep{dunlopetal2002}. Thus, sub-mm observations of powerful radio galaxies over
a wide range of redshifts can offer insight into the formation history of at
least a subset of the most massive elliptical galaxies.

The principle result of this study was that the sub-mm detectability of
powerful radio galaxies was found to be a strong function of redshift, with a
detection rate of $\simeq$ 75\% at $z > 2.5$ compared to only $\simeq $ 15\%
at $z < 2.5$ \citep{rgmnpaper}. Indeed, the average sub-mm luminosity of the
radio galaxies in the sample could be approximated as rising with increasing
redshift with $L \propto (1 + z)^3$ out to $z > 3$.

By definition, all the objects observed in this study contained sufficient 
gas to fuel the activity of their central black holes at the epoch of 
emission. Viewed in the light of this selection bias, the observed strong 
redshift dependence of sub-mm luminosity provides rather convincing evidence 
that the hosts of massive black holes at high redshift are very different to 
the relatively passive elliptical hosts of low-redshift AGN. Specifically, 
irrespective of whether the dust is heated primarily by the UV output of 
young stars or by the AGN itself, these results indicate that the {\it mass} 
of dust (and hence gas) in high-redshift AGN hosts is much greater at $z > 
2.5$. The obvious implication is that by $z \simeq 3$ we are probing an era 
in which a significant fraction of the eventual stellar population of a 
typical massive elliptical has yet to be formed.

Interestingly, a comparison of the redshift distribution of the sub-mm detected
radio galaxies with current best-estimates for the redshift distribution of
sub-mm {\it selected} sources indicates that they are statistically 
indistinguishable \citep{rgmnpaper,sibk2000}. This provides some justification 
for believing that the sub-mm evolution observed for the radio-galaxy 
sample may apply to spheroid formation in general.  However, this then raises 
the obvious counter question of why so few bright sources {\it discovered} via 
sub-mm surveys display any obvious signs of powerful AGN activity?

Motivated by these new observational results we here develop further a simple
model for the dust-enshrouded formation of massive spheroids which we have
previously applied to explore the red envelope of galaxy evolution at
optical-infrared wavelengths \citep{prematuredismissal}. This model is
constructed by combining the stellar population and dust-formation models of
\citet{rauldustmodel, raulssp2002} with a multi-zone chemo-hydrodynamic model
for a collapsing gas spheroid developed by \citet{friacasfr}. The model yields
predictions for the time evolution of gas, dust and the stellar population of
the galaxy as a whole, but also allows us to set constraints on the growth and
fueling of a central black hole by tracking the rate at which gas is deposited
into the central 100 pc of the galaxy. The predictions of this model not only
provide insight into the results of the SCUBA radio-galaxy survey, but also
offer a possible explanation for the relative failure of X-ray surveys to
detect bright SCUBA sources. The model also demonstrates that a correlation
between black-hole and spheroid mass consistent with that seen in low-redshift
galaxies can be produced without invoking a direct causal connection between
black-hole emission and the termination of star formation in the host (as
suggested by \citeauthor{silkrees1998} \citeyear{silkrees1998} and
\citeauthor{fabian1999} \citeyear{fabian1999}).

At first sight, it might appear that this spherically symmetric
calculation of spheroid formation within an isolated and complete dark
matter halo is of dubious relevance in the context of the accepted
paradigm of hierarchical galaxy formation driven by halo mergers.
However, to date, multi-zone chemo-hydrodynamic models of the type
considered here have only been developed in one dimension.  Despite
this technical limitation, we feel it is nevertheless important to
consider the detailed predictions of such a model, especially given
the extremely simplified recipes of star-formation currently adopted
in the construction of semi-analytic models of galaxy
formation\citep[e.g,][]{somervilleprimack1999,benson2001}.  Moreover,
for AGN host galaxies forming at $z \simeq 3$ (i.e. at or near the
peak epoch of quasar activity), the simple model presented here may in
fact offer a good approximation to the average behaviour of the
baryons within the forming host galaxy.  This is because, in such
highly-biased regions within the early Universe, the merging of
collapsed dark matter haloes of mass $\simeq 5 \times 10^{10} \,{\rm
M_{\odot}}$ into a final halo mass $\simeq 10^{13} \,{\rm M_{\odot}}$
is completed very rapidly (see
\sec{connection}).

The structure of this paper is as follows. In \sec{model} we describe the key
components of our theoretical model of a young elliptical galaxy, and
summarize the predicted time evolution of the observable properties of such an
object. The predictions of this model at sub-mm wavelengths are then compared
with the results of the sub-mm study of radio galaxies in
\sec{modeldatacomparison}. In \sec{blackhole} we explore the predicted
fueling and growth of the central black hole, and compare the predictions of
the model with measurements of black-hole masses at low redshift and with the
cosmological evolution of AGN activity. Finally, in \sec{connection} we
discuss the implications of this model for the connection between black-hole
and host-galaxy evolution, focusing on the predicted sub-mm detectability of
AGN, and the X-ray detectability of sub-mm selected sources.  For the
calculation of physical quantities we assume a flat universe with $\Omega_M =
0.3$, $\Omega_{\Lambda} = 0.7$ and $H_0 = 70~{\rm km s^{-1} Mpc^{-1}}$
throughout.

\section{Model of a forming elliptical}
\label{model}

The stellar populations observed in massive elliptical galaxies, 
especially those residing in clusters, appear to be highly coeval, 
consistent with the majority of the stars having formed on a timescale 
$\simeq 1\;$Gyr \citep[e.g.,][]{bowerluceyellis2,bowerluceyellis1}.  
Whether these stars formed in a single region, or in smaller separate 
regions which merged to form the final spheroid, the star formation is 
expected to be vigorous, comprehensive, and obscured by dust - the initial 
gas reservoir being quickly consumed in the star formation process, and
the remnants of the process being blown away by stellar winds and
supernovae \citep[e.g.,][]{zepfandsilk}.

\citet{rauldustmodel} have developed multi-frequency models of galaxies 
formed via such rapid and violent dust-enshrouded starbursts. These models 
predict the time evolution of the ultraviolet-millimeter spectrum of 
a young elliptical galaxy based on a calculation of the star-formation
rate and the resultant chemical evolution of the interstellar medium (ISM).

\subsection{How the Models Work}

There are three key components to the model: (i) the stellar emission from 
the model galaxy, (ii) the re-processing of this emission by dust, and (iii) 
the star-formation law which underpins both these processes.

\subsubsection{Evolution of the Stellar Population}

Given a star-formation law and an initial mass function, the stellar
populations that form in each time step build up to form a complete model
galaxy \citep{raulssp1998, raulssp2002}.  Given a value for metallicity and
the initial stellar mass (governed by the IMF), the evolutionary track of a
star is calculated by the code JMSTAR15 \citep{rauljmstar1,rauljmstar2}, which
includes a proper treatment of the post-main-sequences phases of evolution.
In the absence of dust and gas the integrated emission of the model galaxy at
any time step is simply the sum of the flux from every star in the population.

The model assumes a Salpeter IMF of $\phi(M) \propto M^{-(1+x)}$,
where $\phi(M)$ is the number of stars formed per unit volume with a
mass $M$, and $x=1.35$.  The IMF includes a low-mass cutoff at
$0.1\,{\rm M_{\odot}}$, and is assumed to be constant in both time and
space.

\subsubsection{Dust Model}

The dust model \citep{rauldustmodel} uses a simplified version of the
\citet{draineandlee} dust emission template, and is based on the formalism of
\citet{xuanddezotti}.

For the purposes of calculating the far-infrared emission, the dust is simply 
assumed to lie in a thin shell which surrounds the entire galaxy,  
illuminated by the uniform radiation field generated by the stellar 
population.  This distribution is not realistic.  However, individual
stars and gas clouds cannot be resolved in high-redshift systems.
Given the lack of knowledge concerning the true distributions, this
simple model is no worse than a more complicated model of dust and
star geometry which cannot possibly be justified by the data.

This simple dust distribution is not, however, deemed appropriate for
calculating the reddening of the galaxy light.  Dense molecular clouds are the
birthplace of stellar clusters, where all the massive stars are formed
\citep[e.g.,][]{ladaetal91}.  It takes $1.5\times 10^7$ years for stellar
winds and supernovae to disperse these clouds, which is approximately the
lifetime of the most massive stars
\citep[e.g.,][]{herbig62,elmlada77,blitzshu80}.  The primary source of
extinction of the galaxy light is the dust inside these molecular clouds, not
the dust lying outside the starburst region (which is the assumed geometry for
estimating the far-infrared emission of the galaxy).  Thus, the model assumes
that for the first $1.5\times 10^7$ years of a star's life, the star is {\em
  only} visible in the infrared, with the optical and ultraviolet emission
completely obscured.  After $1.5\times 10^7$ years, the molecular cloud is
assumed to have been dispersed, and the star becomes visible in the optical
and ultraviolet regimes.

\subsubsection{Star-Formation and Chemical Evolution}

The model uses a star-formation law developed by \citet{friacasfr}.  Assuming
spherical symmetry, the galaxy is divided into 100 radial zones.  A 1-D
hydro-dynamical code is used to follow the radial movement of the ISM and
stars, including episodes of gas infall and outflow.  The star-formation rate
has a power-law dependence on the gas density: $SFR(r,t) \propto
\rho^{0.5}$.  The star-formation rate also has a radial dependence on the flow
of the gas.  

Using the resulting distribution of the gas flow, the chemical
evolution is solved by taking into account the lifetimes of stars, and
using modern nucleosynthesis predictions to compute the yields of
stars of different masses. In particular, for stars with masses $0.8 <
M/M_{\odot} < 8$ the yields were taken from \citet{renzini81}, while
for stars with $8 < M/M_{\odot} < 10$ the yields from \citet{hille82}
were assumed.  For SNII progenitors with $10< M/M_{\odot} < 40$ we
adopted the \citet{ww95} yields and extrapolated these for $M > 40
M/M_{\odot}$. Finally, for SN Ia we adopted the yields from
\citet{nomoto84}.

As star formation continues the interstellar gas is heated by Type II
supernovae.  Eventually, the thermal energy of the gas is enough to overcome
the escape velocity, and the remaining gas is expelled from the galaxy.  At
this point, the primary burst of star formation ceases.  Later on, the red
giants of low mass stars and Type I supernovae replenish the gas reservoir and
metal reserves of the ISM, albeit to a limited extent.  There is also a fresh
infall of primordial gas which further dilutes the metallicity of the ISM.
These new gas reserves allow residual star formation to occur.

\subsection{Time Evolution of the Models}

\begin{table}
  \centering
  \begin{tabular}{lc}
      \hline
      Event 			&	Timescale\\
      \hline
      50$\%$ stars formed     	&       0.3$\;$Gyr\\
      98$\%$ stars formed     	&       1.0$\;$Gyr\\
      100$\%$ stars formed    	&       1.7$\;$Gyr\\
      dust mass peaks         	&       0.2$\;$Gyr\\
      metallicity peaks       	&       2.0$\;$Gyr\\
      \hline
  \end{tabular}
\caption{Key stages of formation as predicted by the dusty starburst
  model.}
\label{keystages} 
\end{table}

Friaca \& Terlevich performed their 1-dimensional hydrodynamical/chemico 
calculations for a range of assumed proto-galaxy gas masses, a fact which 
we exploit in Section 4.2. However, for comparison with the properties of 
radio galaxies we adopt a fiducial model which yields an eventual stellar 
mass of $\simeq 10^{12}\,{\rm M_{\odot}}$. The key stages of formation of a 
massive elliptical galaxy, as predicted by this dust-enshrouded starburst
model are summarized in \tab{keystages}.

The primary burst of formation is rigorous and intense, taking only $\sim
0.3\;$Gyr to form the first half of the stellar population.  At its peak, the
star-formation rate is no higher than a few $1000\,{\rm M_{\odot} yr^{-1}}$.
As the primary burst of formation draws to a close, subsequent episodes of
formation occur due to the replenishment of the gas reservoir. This residual
formation is less vigorous, and although $98\%$ of the stellar population is
formed within a Gyr, it takes a further 0.7$\;$Gyr to form the final $2\%$.

One prediction of this model which appears to be a generic prediction of
models of dust-enshrouded star formation is that the dust mass peaks when the
galaxy is approximately half formed. For example, \citet{ee98} and
\citet{fb97} have developed analytical equations which govern the relationship
between gas depletion, star-formation, and dust creation. These equations also
predict a maximum in dust mass half-way through the star formation process.
This is a consequence of competing processes: star formation is required to
create the dust, but star formation also consumes the metal-enriched gas which
houses the dust produced by previous generations of massive stars.  It
transpires that these competing effects balance to produce a maximum in dust
mass approximately half-way through formation of the stellar population.

Even though half of the stars are formed very rapidly, in the first 0.3 Gyr,
it can be seen from Table 1 that star formation persists in the host at some
level for more than a Gyr. As demonstrated by \citet{prematuredismissal}, at
optical-infrared wavelengths this has the important effect of predicting that,
even for a formation redshift $z_f = 5$, the largest values of $R-K$ observed
from an evolving elliptical are produced at $z \simeq 1.5 - 2$. This
prediction counters previous suggestions that pseudo-monolithic models for
elliptical formation can be excluded in favor of hierarchical models simply
due to lack of extremely red objects at high redshift \citep{zepf97}.  In the
current context, this extended tail of star-formation activity leads to the
prediction that a massive elliptical should be detectable at the 1-2 mJy level
at 850${\rm \mu m}$ up to $\sim 2.5$ Gyr after the onset of star-formation activity.

\section{Galaxy Evolution - comparison with radio-galaxy data}
\label{modeldatacomparison}

\subsection{An Example: 4C41.17}

The multi-frequency evolution of galaxy luminosity predicted by the model 
can be used to assess the evolutionary status of well-observed 
massive galaxies at high redshift. As an example of how this can 
work  we here consider two specific predictions of the model. The 
first is the predicted evolution of sub-mm flux density,  and the second is 
the predicted evolution of optical-sub-mm color. The first of these
is obviously an absolute property, dependent not only on the viability
of the model but also on galaxy mass and assumed cosmology. In contrast 
the second is a relative measure of the mass of dust compared to the 
mass of stars which have already formed. While reliable estimates of galaxy
mass at high redshift are not yet achievable, one can at least ask the question
of whether, for reasonable values of assumed mass,  these two predictions 
lead to the same conclusion about the evolutionary status of a given 
object.

The $z = 3.8$ radio galaxy 4C41.17 is one of the best studied of all
objects known at $z > 3$ \citep[e.g.,][]{cmvb90,dhr94,dey97,vbs98}.
Two other properties of 4C41.17 make it particularly useful as a test
of the model predictions. First, as the host of one of the most
powerful known radio sources, it would seem safe to assume that it is
the progenitor of a massive elliptical with an eventual stellar mass
of at least $10^{12}\,{\rm M_{\odot}}$.  Second, its observed optical
emission is known to be dominated by starlight rather than the AGN
\citep{dey97}.

\begin{figure}
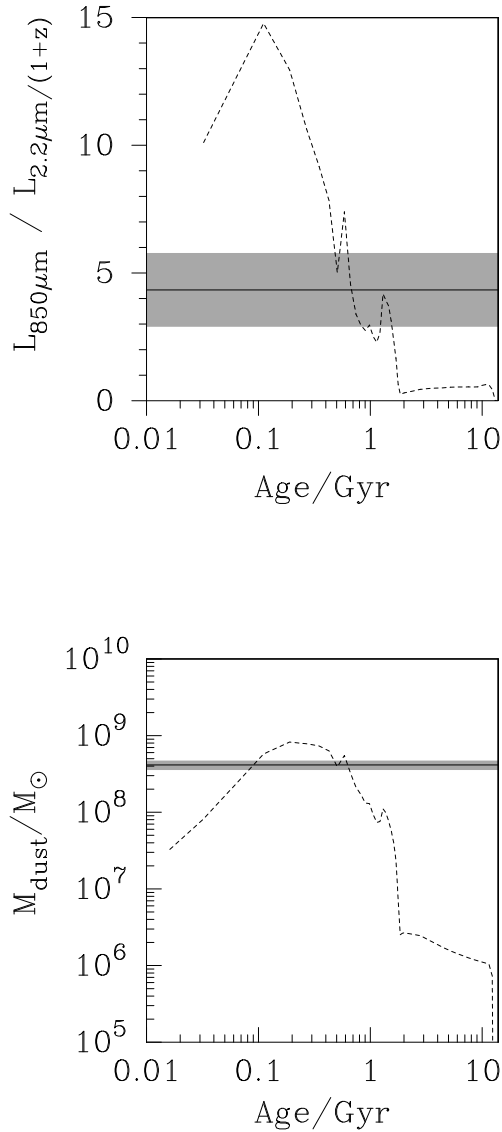

\centering
  \epsfig{file=f1a.eps,width=85mm}
  \epsfig{file=f1b.eps,width=85mm}
\caption{Evolution of \subcolour{} and $M_{dust}$ with time as
  predicted by the dusty starburst model (dashed lines).  The corresponding
  values for 4C41.17 are plotted as horizontal solid lines, with the 
$2\sigma$ error bounds indicated by the shaded region 
\citep{rgmnpaper,vbs98,mythesis}.}
\label{4c41example}
\end{figure}

In \fig{4c41example}, we compare the time evolution of \subcolour{} and 
$M_{dust}$ predicted by the model with the observed values 
for 4C41.17.  This comparison indicates that 4C41.17 is experiencing a 
starburst which is $\sim 0.5-0.8\;$Gyr old, having begun at $z_f\simeq5-6$.  
If the model is a correct description of 4C41.17, then 
$\sim 75\%$ of the final stellar mass of the galaxy has already been formed.

This conclusion is driven primarily by the value of the ratio
\subcolour{} observed for 4C41.17 which, in essence, indicates that
while active star-formation is still continuing at a high level
$\simeq 1000\,{\rm M_{\odot} yr^{-1}}$ the majority of the stellar
population is already in place. But we can see that this conclusion is
also consistent with the observed 850${\rm \mu m}$ flux-density of
4C41.17 provided its total baryonic mass exceeds $\sim 3\times
10^{11}\,{\rm M_{\odot}}$, which seems perfectly reasonable given the
stellar masses of the hosts of low-redshift radio sources \citep{dunlopetal2002}.  Thus, consistent with the original conclusions
of \citet{dhr94}, this model comparison indicates that 4C41.17 is
observed in the final stages of its formative starburst.

If this conclusion is correct, it is worth noting that the model predicts
an object as massive as 4C41.17 would have produced a sub-mm flux density 
as large as $S_{850} \simeq 15-20$ mJy if it had been observed only 0.3-0.6 Gyr
earlier, only 0.2 Gyr after formation (i.e. at $z = 4.5$ for a formation 
redshift $z_f = 5$). However, if observed at this stage in its evolution
the model also predicts that the galaxy would have been much harder to 
detect in the near-infrared, with $K \simeq 21$. Moreover, as discussed
further below, at this stage it seems unlikely that 
4C41.17 would have been capable of powerful radio emission. The inferred
properties of 4C41.17 only $\simeq$ 0.5 Gyr prior to the actual observed
emission epoch are therefore interestingly similar to the observed 
properties of many of the sub-mm selected sources found in the blank-field 
SCUBA surveys \citep{dunlop2001}. This point is pursued 
further in section 5.

\subsection{Redshift dependence of sub-mm luminosity}

\begin{figure}
\centering
  \epsfig{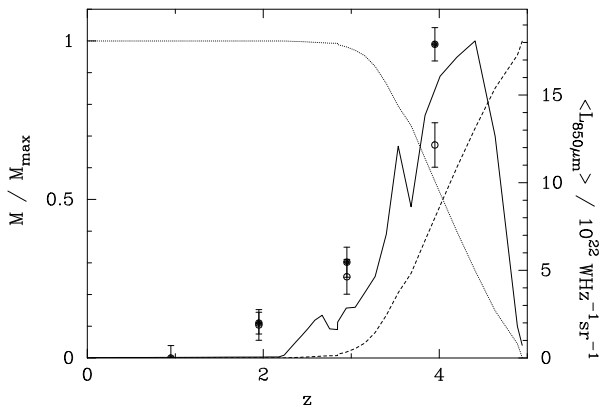}
\caption{Redshift evolution of a massive spheroid as predicted
  by the model for an assumed formation redshift $z_f=5.0$.  The dotted line
  represents the fractional stellar mass, the dashed line represents the
  fractional gas mass, and the solid line shows the dust mass normalized to
  its maximum value.  The data points give the weighted-mean 850-\micron{}
  luminosity, $<$\luminsub$>$, of the radio galaxy sample binned in redshift
  \citep{rgmnpaper}.  The solid circles indicate the entire dataset; the open
  circles indicate a subset chosen to remove any radio-luminosity bias
  \citep[see ][]{rgmnpaper}.}
\label{modelmassesz}
\end{figure}

Assuming a formation redshift of $z_f=5.0$ as inferred above for 4C41.17, we plot in
\Fig{modelmassesz} the redshift evolution of the gas mass, dust mass, and
stellar mass predicted by the model.

The formation of the spheroid lasts approximately 1 Gyr and is $\simeq$ 95\%
complete by $z \simeq 2.5$. While it would seem unreasonable to ascribe a
single formation redshift $z_f = 5$ to all powerful radio galaxies, it is
nevertheless interesting that $z \simeq 2.5$ is also the redshift below which
radio galaxies generally appear as relaxed spheroids in infrared images, and
above which they generally appear to be more complex, extended and clumpy
sources in the rest-frame optical \citep{vbs98,pent2001}.
Moreover, as also illustrated in \Fig{modelmassesz}, the predicted 
redshift evolution of sub-mm luminosity is undeniably 
reminiscent of the redshift dependence of mean \luminsub{} 
for the full radio-galaxy sample observed by \citet{rgmnpaper}.

In summary, the observed sub-mm - optical properties of 
high-redshift radio galaxies are, at least to first order, consistent
with the model predictions for an assumed typical formation redshift 
$z_f \simeq 5$.

\subsection{Redshift dependence of optical luminosity}

\begin{figure}
\centering
  \epsfig{file=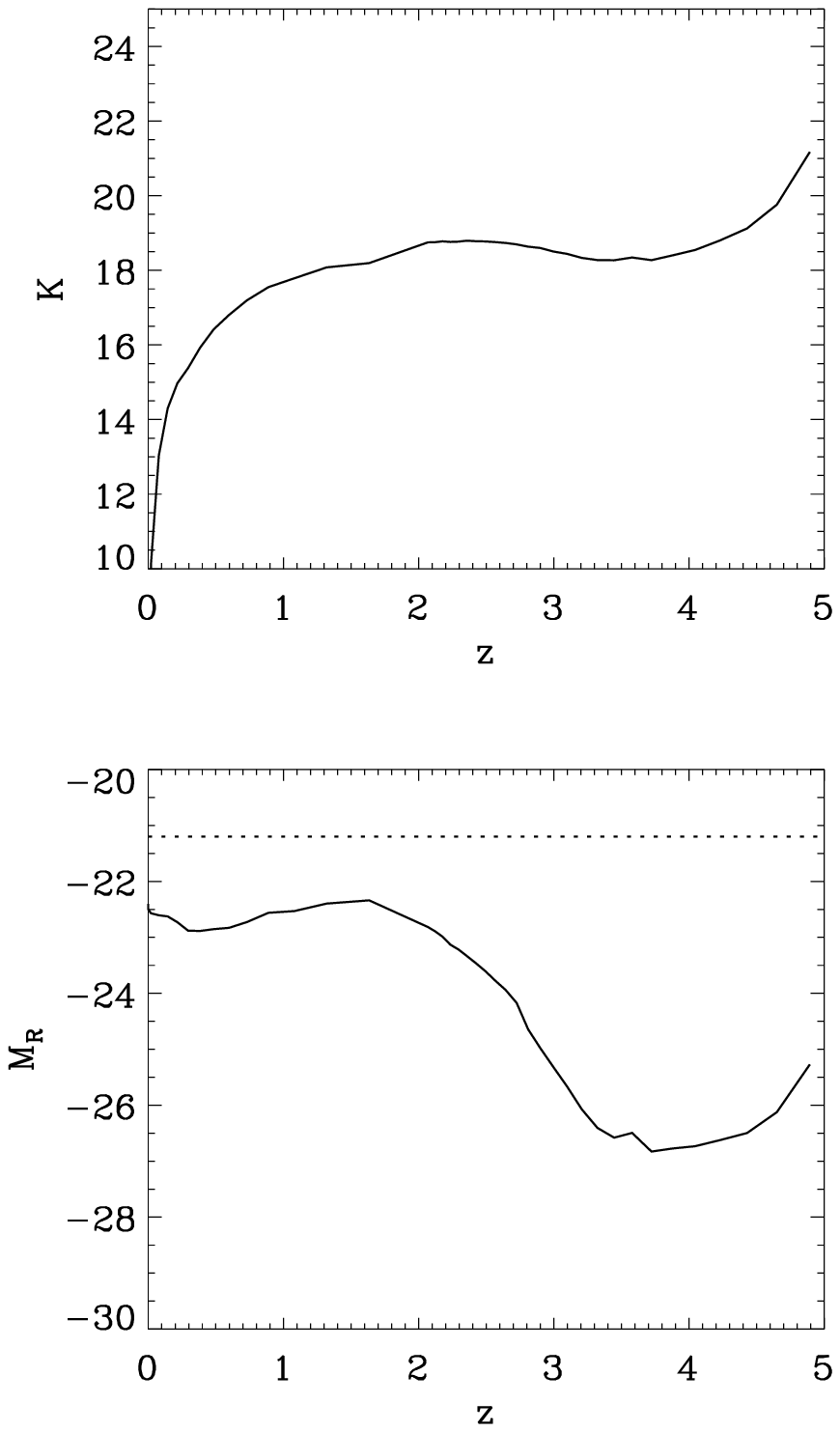,,angle=0,width=85mm}
\caption{
Top panel: predicted evolution of 
apparent $K$-band magnitude as a function of redshift. Bottom
panel:  predicted evolution of
absolute $R$-band magnitude as a function of redshift.
The dotted line shows the corresponding absolute $R$ magnitude for a
present-day $L^{\star}$ galaxy.}
\label{rkz}
\end{figure}

Finally, we plot in \Fig{rkz} the redshift evolution of optical luminosity,
both in terms of absolute $R$-band magnitude, and in terms of observed 
$K$ magnitude. As can be seen from the lower plot, the $z = 0$ $R$-band 
absolute magnitude is 1.4 magnitudes brighter than $M_R^{\star}$, consistent
with observations which indicate that radio galaxies are typically $3-4$
times more luminous than $L^{\star}$ galaxies. 

\section{Black Hole Evolution}
\label{blackhole}

Having established that the model provides a plausible description of the 
evolution of massive ellipticals as traced by radio galaxies, 
we now explore the predictions of the model for central black-hole growth, 
and AGN output as a function of time/redshift.

\subsection{Black Hole Growth}

As detailed in \citet{friacasfr} one output of the multi-zone model
is the amount of gas delivered per unit time to the central regions 
of the galaxy ($r < 100$ pc).

\begin{figure}
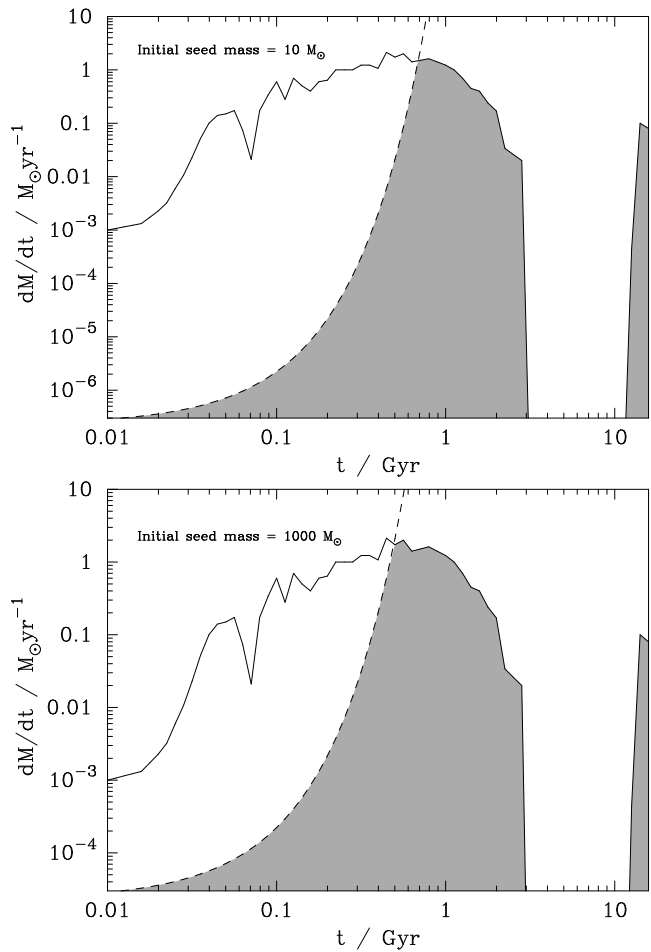

\centering
  \epsfig{file=f4a.eps,angle=-90,width=85mm}
  \epsfig{file=f4b.eps,angle=-90,width=85mm}
\caption{Black-hole fueling rate as a function of time,
  assuming initial black-hole seed masses of $10\,{\rm M_{\odot}}$ and
  $1000\,{\rm M_{\odot}}$. The solid line depicts the rate at which gas is
  deposited into the central 100 pc of the model galaxy.  The dashed line
  represents the rate of gas consumption by the black hole assuming that this
  rate cannot be greater than that dictated by the Eddington limit for an
  assumed radiation efficiency of 10$\%$. The shaded area indicates that
  Eddington-limited accretion persists until $\sim 0.5 - 0.7\,$Gyr, beyond which
  the central cooling flow cannot sustain Eddington-limited growth. Most of
  the final black-hole mass is then accreted at sub-Eddington rates over the
  subsequent $\simeq 1$ Gyr.}
\label{bhfueling}
\end{figure}

To calculate the growth of the central black hole we assume that all of this
fuel is available for consumption by the central black hole \citep[true in the absence of any other limiting factor,][]{burkertsilk2001}. However, as
illustrated in \Fig{bhfueling} (for our fiducial model $M_{gal} =
10^{12}\,{\rm M_{\odot}}$) in practice for most of the first Gyr, the growth of the
central black hole is limited not by this fuel supply, but by the Eddington
limit \citep[following the prescription of][]{smallblandford92}, unless a very large seed mass ($\simeq 100000\,{\rm M_{\odot}}$) is adopted.
\Fig{bhfueling}a shows how black-hole mass consumption compares with central
fuel delivery as a function of time for an assumed seed mass of
10$\,{\rm M_{\odot}}$ and an assumed black-hole radiative efficiency
of 10$\%$, while \Fig{bhfueling} shows the effect of increasing the
seed mass to 1000$\,{\rm M_{\odot}}$. We consider $10-1000\,{\rm
M_{\odot}}$ to represent a reasonable range for assumed seed mass
because observationally no stars with masses larger than $\simeq
200\,{\rm M_{\odot}}$ are known \citep{figer98}, and because theoretical
models indicate that the first generation of population III stars may
have produced black holes of mass $\simeq 100\,{\rm M_{\odot}}$
\citep{fryerinpress,madaurees2001}.

\begin{figure}
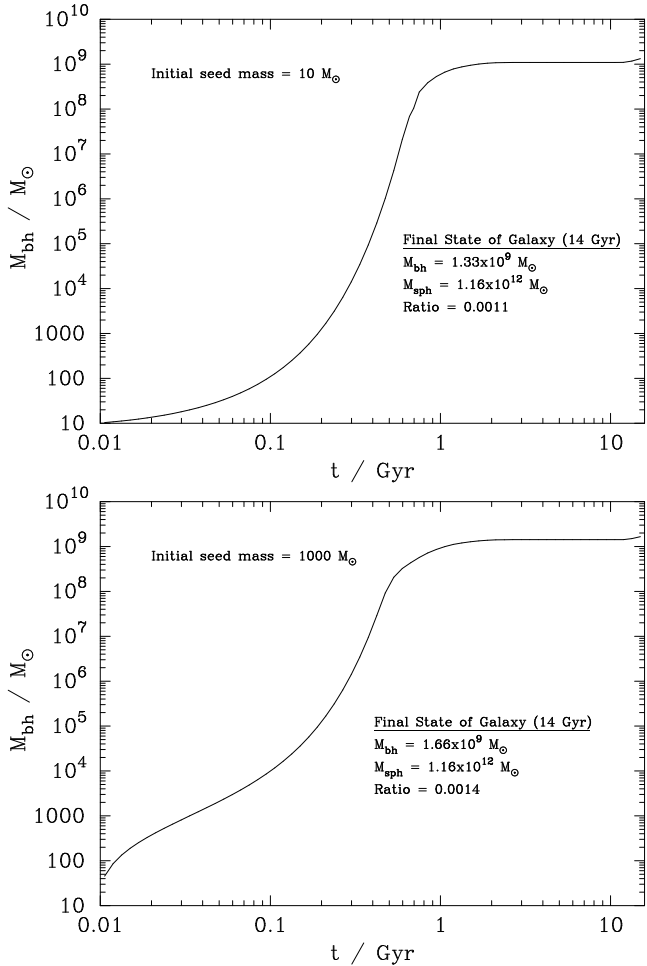

\centering
  \epsfig{file=f5a.eps,angle=-90,width=85mm}
  \epsfig{file=f5b.eps,angle=-90,width=85mm}
\caption{As \fig{bhfueling}, but this time plotting the the
  logarithm of black-hole mass as a function of time, for both alternative
  assumed seed masses.}
\label{bhgrowth}
\end{figure}

Figures \ref{bhgrowth}a and \ref{bhgrowth}b show how black-hole mass
grows as a function of time for both assumed seed masses. In the first
case the final black-hole mass is $1.3 \times 10^9\,{\rm M_{\odot}}$,
while in the second case it is $1.6 \times 10^9\,{\rm
M_{\odot}}$. Thus, varying the seed mass by a factor of 100 has only a
very minor effect on the final mass of the black hole.  Indeed, even
if it is assumed that all the mass deposited into the central 100 pc
is ultimately consumed by the black hole (i.e. including that fraction
of gas which is delivered, but cannot be consumed during the
Eddington-limited accretion phase) the final mass is still not very
different, at $2.0 \times 10^9 \,{\rm M_{\odot}}$.

Inspection of Figures \ref{bhfueling} and \ref{bhgrowth} shows that
the black hole grows and radiates at the Eddington limit for the first
$\simeq 0.5 - 0.7$ Gyr (doubling in mass every $3 \times 10^7$ yrs)
but that beyond that point, when the black hole has grown to a mass of
a few times $10^8\,{\rm M_{\odot}}$, the fuel delivery is unable to
sustain further Eddington-limited fueling/growth. However, the black
hole continues to grow to its final mass in excess of $10^9\,{\rm
M_{\odot}}$ for a further Gyr or so, through sub-Eddington accretion
before the gas supply is essentially switched off 3 Gyr after
formation.

Although this paper does not pretend to place any new constraints on
our understanding of black-hole accretion mechanisms, it is
nevertheless important to ask whether the accretion rates produced by
this model are consistent with current understanding of the rate at
which a black hole can consume matter from an accretion disc.  In
particular, one might question whether the black hole is capable of
consuming the gas delivered to it at the peak rate which approaches $2
\,{\rm M_{\odot} yr^{-1}}$ (see \fig{bhfueling}).  In fact, for a
massive spheriod (with central velocity dispersion $\sigma_{sph}
\simeq 200 \,{\rm km s^{-1}}$) the calculations of
\citet{burkertsilk2001} indicate that accretion disc viscosity can be
expected to limit the mass consumption rate of a super-massive black
hole at the centre of a forming spheriod to $\simeq 2-20 \,{\rm
M_{\odot} yr^{-1}}$ for a critical Reynolds number $R_{cr}$ in the
range $1000-100$ \citep{duschl2000}.  Thus, our current knowledge of
accretion disc physics does not appear to conflict with our assumption
that the black-hole can consume all the gas delivered to the central
region of the galaxy at a peak rate which, in this particular model,
approaches $2 \,{\rm M_{\odot} yr^{-1}}$.

Several features of these results deserve comment. First, due to the
assumption of Eddington-limited growth, the spheroid does not contain
a black hole of sufficient mass to host a quasar (i.e. $M_{bh} >
10^8\,{\rm M_{\odot}}$; Dunlop et al. 2002) until $0.5-0.7$
Gyr after commencement of star-formation. This predicted delay is
still consistent with the discovery of quasars at $z \simeq 6$ in our
favored cosmology ($t_{universe} = 0.9$~Gyr at $z = 6$), and can only
be significantly accelerated by the assumption of either a very large
seed mass or possibly runaway merging of black holes as suggested by
\citet{ebisu2001}.

Second, the final black-hole mass results in a black-hole:spheroid
mass ratio of $0.0011-0.0014$ (or 0.0017 if all available gas is
consumed by the black hole), in very good agreement with observational
results at low redshift \citep{kormendy2001,mf2001}.  Note that this
ratio arises not from invoking the assumption that the black-hole
itself is responsible for regulating and terminating star-formation in
the host
\citep{silkrees1998,fabian1999}, but rather because the gas dynamics and
star-formation in the forming spheroid has not allowed
Eddington-limited exponential black-hole growth to continue beyond the
first Gyr, and ultimately terminates significant black-hole fueling
after the first $2-3$ Gyrs (we note that Burkert \& Silk 2001 also
appeal to star formation to limit excessive black-hole growth, but in
their model the star-formation in question is specifically assumed to
be in the outer accretion disc).
 
Third, there is a window lasting $\simeq 1$ Gyr where the black hole has 
reached a mass sufficient to be capable of producing a powerful AGN and 
during which it is still receiving fuel at a sufficient rate 
($\simeq 1\,{\rm M_{\odot} yr^{-1}}$) to do so. The bulk of the mass of the
black-hole is actually accreted at sub-Eddington rates during this 
`AGN epoch'.

Fourth, the epoch of peak sub-mm luminosity precedes peak AGN luminosity
by $0.3 - 0.5$ Gyr, due to the fact that sub-mm output peaks half-way through
spheroid formation. Indeed, dependent on assumed seed mass, the black-hole 
mass at the time of peak sub-mm emission is only 
$10^4 - 10^6\,{\rm M_{\odot}}$, offering an interesting alternative 
explanation (c.f. pure obscuration) for the failure to detect bright SCUBA 
sources at X-ray wavelengths (see \sec{connection}).

\subsection{The black-hole:spheroid mass relation}

\begin{figure}
\centering
  \epsfig{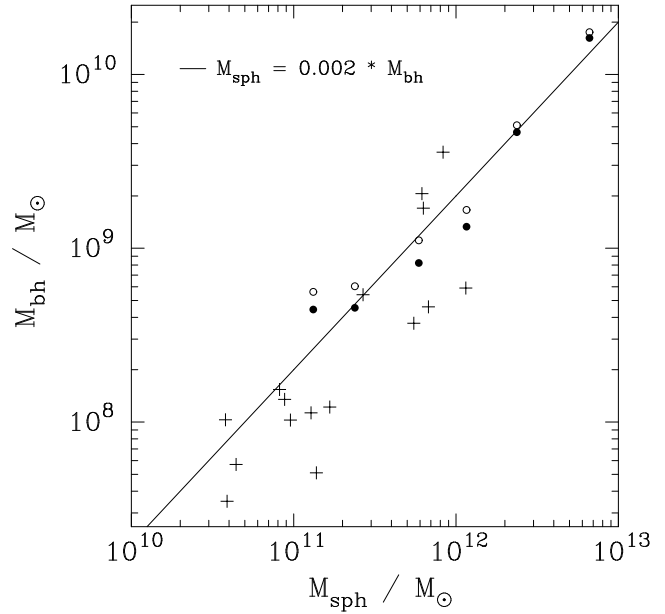}
\caption{Final black-hole mass plotted against final spheroid
  mass, for the range of galaxy masses explored by \citet{friacasfr}.  The
  solid and open circles indicate the results assuming initial seed masses of
  10 and 1000$\,{\rm M_{\odot}}$ respectively.  The crosses indicate the 17
  low-redshift galaxies which currently possess the most reliable estimate of
  black hole mass.  The values of $M_{sph}$ plotted for these objects are the
  revised values given by \citet{rossjim2002} from their
  reanalysis of galaxy bulge luminosity.  Both this dataset and the model
  prediction are clearly consistent with a linear relation of the form $M_{bh}
  = k\; M_{sph}$.  The solid line indicates the best fitting relation to the
  model prediction, with a normalization $k = 0.002$.}
\label{masscomparison}
\end{figure}

Given the rather successful prediction of our fiducial model that a
host spheroid with mass $M_{sph} \simeq 10^{12}\,{\rm M_{\odot}}$
should house a black-hole of mass $M_{bh} \simeq 1.5 \times
10^{9}\,{\rm M_{\odot}}$, it is obviously of interest to undertake the
same calculation for the range of spheroid masses originally modeled
by \citet{friacasfr}.  The results of this calculation are presented
in \Fig{masscomparison}, where predicted black-hole mass (for seed
masses of 10 and $1000\,{\rm M_{\odot}}$) is plotted against spheroid
mass for five different galaxy masses. The best-fitting relation to
these predicted data has a slope of 0.002, consistent with current
estimates of the normalization of the $M_{bh}:M_{sph}$ relation
\citep{kormendy2001,mf2001}.

The fact that this relation can be successfully predicted without any
need to appeal to a causal connection between black-hole radiation and
gas ejection is perhaps unexpected, but obviously interesting. It
suggests that this relation may simply come about because the amount
of gas delivered to the central regions of a forming spheroid is
roughly proportional to the total baryonic mass of the spheroid, with
the rate and duration of gas delivery to the central region being
limited by the star-formation process.

\subsection{AGN Activity}

Detailed observations of low-redshift AGN indicate that production of
a QSO with $M_V < -23$ requires a black-hole mass $M_{bh} > 10^8\,{\rm
M_{\odot}}$ while production of a radio-loud source with $P_{2.7GHz} >
10^{25}~{\rm W Hz^{-1} sr^{-1}}$ requires $M_{bh} > 10^9\,{\rm
M_{\odot}}$ \citep{dunlopetal2002,laor98,rossjim2002}.  Although
current constraints are poor, existing data on quasars at high
redshift certainly do not contradict the hypothesis that these mass
thresholds are physically significant and still apply even at $z
\simeq 2$ \citep{kukula2001}. If we now apply these mass thresholds to
the model predictions shown in
\fig{bhgrowth}, we can see that the model predicts that quasar-level optical
activity cannot be produced until $\simeq 0.5$ Gyr after commencement of star
formation in the host, while production of a powerful radio source is delayed
by $\simeq 1$~Gyr.

However, the model predictions potentially extend beyond the basic prediction
of such time delays. In particular, reference to Figures \ref{bhfueling} and
\ref{bhgrowth} shows that, until the black hole passes the aforementioned mass
threshold of $\simeq 10^8\,{\rm M_{\odot}}$, the rate of gas delivery into the
central regions of the galaxy by the cooling flow is well in excess of that
required to sustain Eddington-limited black-hole growth. As discussed by
\citet{fabian1999}, in this situation it might be reasonably assumed that all
optical emission, and a substantial fraction of soft X-ray emission from the
growing black-hole is extinguished and that in its rapid growth phase the
black hole may be only readily observable at hard X-ray wavelengths.

The model thus makes a number of predictions about the 
observability of high-redshift AGN at different wavelengths which can be 
compared with the results of current and future observations. These can be
summarized as follows.

\begin{enumerate}
  
\item{Assuming that, due to bias, massive spheroids such as that represented by
    our fiducial model are the first to collapse, the first AGN observable at
    optical wavelengths should be rather luminous QSOs, emerging $\simeq 0.5$
    Gyr after the onset of massive star formation in the galaxy halo. This is
    because, by the time the central black hole ceases to be smothered by the
    central cooling flow and enters the regime of sub-Eddington accretion, it
    has already grown to a mass $M_{bh} > 10^8\,{\rm M_{\odot}}$. This
    prediction is at least consistent with the relative success of bright
    quasar surveys in locating high-redshift QSOs \citep{fan1}.}

\item{For a formation redshift $z = 5$, a powerful QSO should therefore emerge
    at $z \simeq 3.5$, simply because Eddington-limited growth of the central
    black hole from a seed mass $< 1000\,{\rm M_{\odot}}$ requires $\simeq 0.5$
    Gyr to produce a black hole of the required mass $> 10^8\,{\rm M_{\odot}}$.
    For any formation redshift, luminous QSOs should not exist at redshifts
    significantly in excess of $z = 7$.  The current quasar redshift record
    stands at $z = 6.28$ \citep{fan3}}. 

\item{For a formation redshift $z = 5$, production of a powerful radio source
    is predicted to be delayed until $z \simeq 3$, based once again on
    Eddington-limited growth and adoption of a minimum black-hole mass
    threshold of $10^9\,{\rm M_{\odot}}$.  For any formation redshift, powerful
    radio sources should not be found at redshifts significantly in excess of
    $z = 5.5$.  The current radio-source redshift record stands at $z = 5.19$
    \citep{vanB1999}.}

\item{Conversely, the predicted $\simeq0.5$-Gyr time lag between the 
     onset of massive 
     star-formation activity and the appearance of a central QSO in our model
     can be used to infer the first epoch of massive star-formation from 
     the redshifts of the most distant QSOs. Given the current QSO redshift
     record of $z = 6.28$, our model would predict that massive 
     star-formation commenced in its host galaxy at $z \simeq 10 -12$.}
 
\item{The very highest redshift AGN should only be visible in hard X-rays, as
    they rapidly approach a mass $\simeq 10^8\,{\rm M_{\odot}}$ via
    Eddington-limited accretion fueled by a cooling flow delivering gas at
    super-Eddington rates to the central regions of the forming galaxy.}

\end{enumerate}

\section{The Starburst-AGN connection}
\label{connection}

\begin{figure}
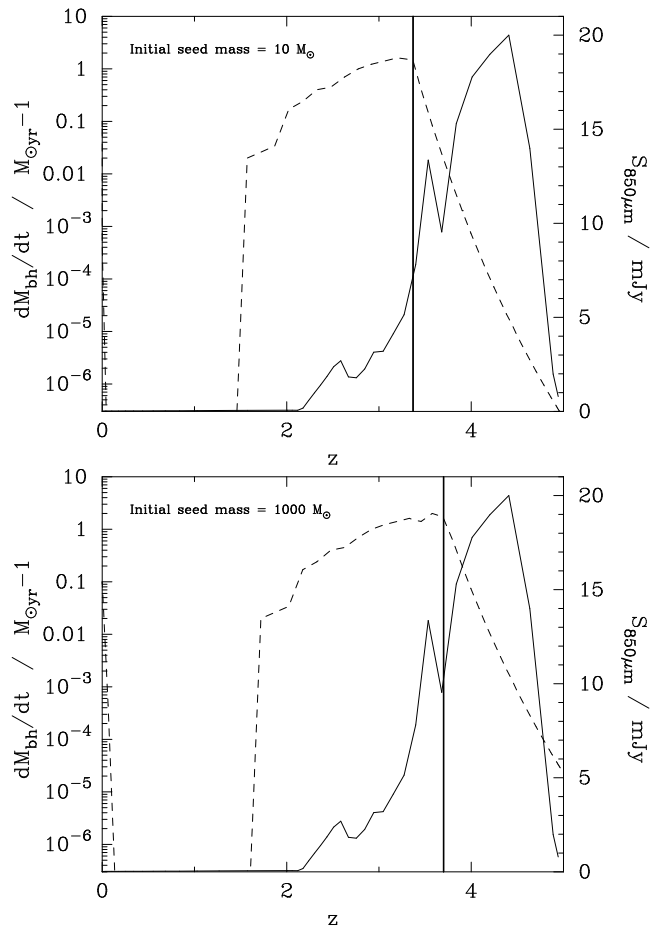

\centering
  \epsfig{file=f7a.eps,angle=-90,width=85mm}
  \epsfig{file=f7b.eps,angle=-90,width=85mm}
\caption{Predicted redshift evolution of the sub-mm output of
  a massive elliptical galaxy formed at $z_f = 5$ compared with the
  predicted redshift evolution of the bolometric output of its central
  AGN, again shown for assumed black-hole seed masses of 10 and
  1000$\,{\rm M_{\odot}}$ respectively.  The dashed line (left-hand
  axis) displays the logarithm of AGN activity, in terms of the
  fueling rate of the central black hole. The thick solid line marks
  the redshift at which the black-hole growth becomes sub-Eddington.
  The thin solid line (right-hand axis) shows the sub-mm output of the
  host, in terms of predicted observed 850\,\micron{} flux
  density. This plot demonstrates that for $z_f = 5$ and the
  currently-favored cosmological model, a massive forming spheroid can
  achieve peak AGN output at $z \simeq 3.5$, while still being
  detectable in the sub-mm with $S_{850} \simeq 10$~mJy, thus
  reproducing almost exactly the observed properties of the $z = 3.8$
  radio galaxy 4C41.17.}
\label{comparison}
\end{figure}

\begin{figure}
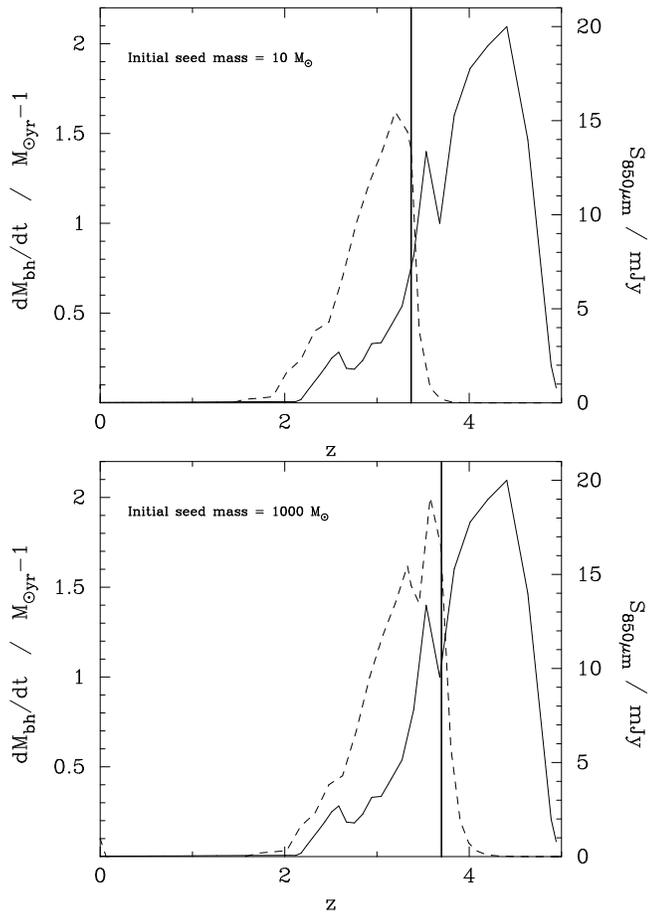

\centering
  \epsfig{file=f8a.eps,angle=-90,width=85mm} 
  \epsfig{file=f8b.eps,angle=-90,width=85mm}
\caption{As \fig{comparison}, but with the AGN activity
  displayed on a linear scale, to better illustrate the rapid rise and demise
  of AGN activity as star formation in the host approaches completion.}
\label{linearcomparison}
\end{figure}

Finally we focus on the issue of the connection between host-galaxy and
black-hole evolution. In \fig{comparison} and \fig{linearcomparison} we plot
together the redshift dependence of AGN output as inferred from Figures
\ref{bhfueling} and \ref{bhgrowth}, alongside the evolution of host sub-mm
luminosity, again assuming a formation redshift $z_f = 5$.

This combined diagram is instructive because it allows one to predict the
properties of objects uncovered at different wavelengths as a function of
flux-density and redshift. We emphasize that although \fig{comparison}
represents the predictions of the specific model considered here, most of the
resulting observable consequences apply to any quasi-monolithic model of
elliptical formation in which star-formation activity peaks in a small
fraction of a Gyr, while the rate of black-hole growth is dictated by
Eddington-limited accretion starting from a modest seed mass.

First, it can be seen that the predicted duty cycle of quasar activity at high
redshift is considerably longer ($\simeq 1$ Gyr) than that which is generally
inferred from (presumably retriggered) activity at low redshift, with
excessive black-hole masses being avoided by the fact that most of the
accretion during this primary, long-lived active phase is sub-Eddington.
Consequently it is hard to escape the prediction that, for a reasonable range
of formation redshifts, a substantial fraction of the present-day massive
elliptical population should be active at the same time around $z = 2-3$. In
fact, provided that one assumes at least half of optical QSOs are obscured by
orientation effects, the comoving number density of quasars at $z \simeq 2.5$
is $\simeq 1 \times 10^{-5} {\rm Mpc^{-3}}$, comparable to the present-day
number density of ellipticals with $L > 2-3 L^{\star}$ \citep{kochanek2001}.

Second, it is clear that sub-mm observations of powerful AGN are unlikely to
uncover the most luminous sub-mm sources, but should be capable of yielding
detections for those objects observed in the overlap period when the
black-hole has grown to large mass but sub-mm luminosity has not yet declined
to undetectably low levels. Statistically, this is clearly most likely to be
the case for the highest redshift AGN, and we would therefore predict that the
sub-mm detectability of powerful AGN should be a strong function of redshift,
approaching unity at $z > 3$. As already discussed this does indeed appear to
be the case for powerful radio galaxies \citep{rgmnpaper}, and we would
predict that the sub-mm detectability of optical and X-ray selected QSOs
should be a similarly strong function of redshift.

Third, it can be seen from \fig{comparison} that, most strikingly for the
assumed seed mass of 1000$\,{\rm M_{\odot}}$, the formation redshift of 4C41.17
inferred from its near-IR to sub-mm color in section 3, does indeed result in
maximum AGN output at its observed redshift of $z = 3.8$ as well as a sub-mm
luminosity at this redshift in excellent accord with the observed value of
$S_{850} = 12\pm 0.9$ mJy \citep{rgmnpaper}.  Thus the model can successfully
account for the basic observed properties of this well-studied object at the
observed epoch.  Given that 4C41.17 appears to be one of the most massive
objects in the universe, \fig{comparison} also indicates that the most
luminous sub-mm sources are unlikely to exceed the sub-mm output of 4C41.17 by
more than a factor $\simeq 2$, and thus sub-mm surveys are unlikely to
uncover essentially any sources brighter than $S_{850} \simeq 20$ mJy.
   
Finally, we would expect such extreme SCUBA sources, and indeed the majority
of sub-mm sources uncovered by the current relatively shallow sub-mm surveys
to be biased toward the epoch of peak sub-mm output. As illustrated in
\fig{comparison} and \fig{linearcomparison} at this stage (which occurs in our
model only $\simeq 0.2$ Gyr after the onset of star-formation), the
growing black hole is likely not only to be heavily obscured, but to
have a black-hole mass (and AGN output) $10^3 - 10^5$ times smaller
than that of luminous quasars.  This means that, regardless of their
precise redshift distribution, the fraction of bright sub-mm detected
sources which should also prove detectable at X-ray wavelengths with
current facilities is small. This prediction is at least qualitatively
consistent with current results which indicate that only $\simeq 10\%$
of bright sub-mm sources show any detectable signs of AGN activity at
any wavelength \citep{fabianxray2000,omarxray2002,bargerxray2001}.  We
emphasize that while the growing black-hole at this stage may well be
heavily obscured, the model presented here predicts that SCUBA sources
are unlikely to be detectable with current Chandra surveys primarily
because the black hole is simply not big enough, rather than by
insisting that the sources are Compton thick with $<1\%$ scattering
\citep{fabian1999}; even for unobscured AGN the deep Chandra
observations presented by \citet{omarxray2002} are incapable of
detecting a black-hole less massive than $\simeq 10^7\,{\rm M_{\odot}}$ at
$z \simeq 3$.  The model therefore also predicts that AGN heating
makes a negligible contribution to the luminosity of high-redshift
SCUBA sources.

In a future paper we intend to explore the effect of applying our
model to track the behavior of the baryons within a CDM merger tree
leading to a massive ($10^{13}~{\rm M_{\odot}}$) collapsed dark matter
halo at $z \simeq 3$.  However, for now we simply note that the simple
model presented here may well offer a good approximation to the
average behavior of the baryons within such a merger tree. This is
because, for such highly-biased regions within the high-redshift
universe, the merging of collapsed dark matter haloes of mass $\simeq
5 \times 10^{10}\,{\rm M_{\odot}}$ into a final halo of mass $\simeq
10^{13}\,{\rm M_{\odot}}$ is completed in a timescale $< 1$~Gyr,
comparable to the timescale of the star-formation and black-hole
formation processes described in this paper.

\section*{ACKNOWLEDGMENTS}
James Dunlop acknowledges the enhanced research time afforded by the award 
of a PPARC Senior Fellowship. Ross McLure acknowledges the support of a PPARC
Fellowship, while Raul Jimenez acknowledges partial support from a PPARC
Advanced Fellowship.

%\bibliographystyle{mn}
%\bibliography{mn_jour,../bibliography/horatio}

\begin{thebibliography}{65}
\expandafter\ifx\csname natexlab\endcsname\relax\def\natexlab#1{#1}\fi

\bibitem[{{Almaini} {et~al.}(2002){Almaini}, {Dunlop}, {Lawrence}, {Blain},
  {Manners}, {Johnson}, {Willott}, {Perez-Fournon}, {Gonzales-Solares},
  {Oliver}, {Ciliegi}, {Serjeant}, \& {Rowan-Robinson}}]{omarxray2002}
{Almaini} O. et al., 2002, MNRAS, in press,
  astro-ph/0108400

\bibitem[{{Archibald}(1999)}]{mythesis}
{Archibald} E.~N., 1999, PhD thesis, University of Edinburgh

\bibitem[{{Archibald} {et~al.}(2001){Archibald}, {Dunlop}, {Hughes},
  {Rawlings}, {Eales}, \& {Ivison}}]{rgmnpaper}
{Archibald} E.~N., {Dunlop} J.~S., {Hughes} D.~H., {Rawlings} S., {Eales}
  S.~A., {Ivison} R.~J., 2001, \mnras, 323, 417

\bibitem[{{Barger} {et~al.}(2001){Barger}, {Cowie}, {Mushotzky}, \&
  {Richards}}]{bargerxray2001}
{Barger} A.~J., {Cowie} L.~L., {Mushotzky} R.~F., {Richards} E.~A., 2001, \aj,
  121, 662

\bibitem[{{Benson} {et~al.}(2001){Benson}, {Frenk}, {Baugh}, {Cole}, \&
  {Lacey}}]{benson2001}
{Benson} A.~J., {Frenk} C.~S., {Baugh} C.~M., {Cole} S., {Lacey} C.~G., 2001,
  MNRAS, 327, 1041

\bibitem[{{Blitz} \& {Shu}(1980)}]{blitzshu80}
{Blitz} L., {Shu} F.~H., 1980, \apj, 238, 148

\bibitem[{{Bower} {et~al.}(1992{\natexlab{a}}){Bower}, {Lucey}, \&
  {Ellis}}]{bowerluceyellis2}
{Bower} R.~G., {Lucey} J.~R., {Ellis} R.~S., 1992{\natexlab{a}}, \mnras, 254,
  601

\bibitem[{{Bower} {et~al.}(1992{\natexlab{b}}){Bower}, {Lucey}, \&
  {Ellis}}]{bowerluceyellis1}
---, 1992{\natexlab{b}}, \mnras, 254, 589

\bibitem[{{Boyle} \& {Terlevich}(1998)}]{BoyleTerlevich1998}
{Boyle} B.~J., {Terlevich} R.~J., 1998, \mnras, 293, L49

\bibitem[{{Burkert} \& {Silk}(2001)}]{burkertsilk2001}
{Burkert} A., {Silk} J., 2001, \apjl, 554, L151

\bibitem[{{Chambers} {et~al.}(1990){Chambers}, {Miley}, \& {van
  Breugel}}]{cmvb90}
{Chambers} K.~C., {Miley} G.~K., {van Breugel} W. J.~M., 1990, \apj, 363, 21

\bibitem[{{Dey} {et~al.}(1997){Dey}, {van Breugel}, {Vacca}, \&
  {Antonucci}}]{dey97}
{Dey} A., {van Breugel} W., {Vacca} W.~D., {Antonucci} R., 1997, \apj, 490, 698

\bibitem[{{Draine} \& {Lee}(1984)}]{draineandlee}
{Draine} B.~T., {Lee} H.~M., 1984, \apj, 285, 89

\bibitem[{{Dunlop}(1997)}]{dunlop1997}
{Dunlop} J.~S., 1997, in Observational Cosmology with the New Radio Surveys,
  {Bremer} M., et~al., eds., Kluwer, p. 157

\bibitem[{{Dunlop}(2001)}]{dunlop2001}
---, 2001, in UMass/INAOE conference proceedings on `Deep millimeter surveys',
  {Lowenthal} J., {Hughes} D.~H., eds., World Scientific, p.~11

\bibitem[{{Dunlop} {et~al.}(1994){Dunlop}, {Hughes}, {Rawlings}, {Eales}, \&
  {Ward}}]{dhr94}
{Dunlop} J.~S., {Hughes} D.~H., {Rawlings} S., {Eales} S.~A., {Ward} M.~J.,
  1994, \nat, 370, 347

\bibitem[{{Dunlop} {et~al.}(2002){Dunlop}, {McLure}, {Kukula}, {Baum}, {O'Dea},
  \& {Hughes}}]{dunlopetal2002}
{Dunlop} J.~S., {McLure} R.~J., {Kukula} M.~J., {Baum} S.~A., {O'Dea} C.~P.,
  {Hughes} D.~H., 2002, MNRAS, in press, astro-ph/0108397

\bibitem[{{Duschl} {et~al.}(2000){Duschl}, {Strittmatter}, \&
  {Biermann}}]{duschl2000}
{Duschl} W.~J., {Strittmatter} P.~A., {Biermann} P.~L., 2000, A\&A, 357, 1123

\bibitem[{{Eales} {et~al.}(1999){Eales}, {Lilly}, {Gear}, {Dunne}, {Bond},
  {Hammer}, {Le F{\` e}vre}, \& {Crampton}}]{ealessurvey1999}
{Eales} S., {Lilly} S., {Gear} W., {Dunne} L., {Bond} J.~R., {Hammer} F., {Le
  F{\` e}vre} O., {Crampton} D., 1999, \apj, 515, 518

\bibitem[{{Ebisuzaki} {et~al.}(2001){Ebisuzaki}, {Makino}, {Go Tsuru},
  {Funato}, {Zwart}, {Hut}, {McMillan}, {Matsushita}, {Matsumoto}, \&
  {Kawabe}}]{ebisu2001}
{Ebisuzaki} T. et al., 2001, ApJL, 562, L19

\bibitem[{{Edmunds} \& {Eales}(1998)}]{ee98}
{Edmunds} M.~G., {Eales} S.~A., 1998, \mnras, 299, L29

\bibitem[{{Elmegreen} \& {Lada}(1977)}]{elmlada77}
{Elmegreen} B.~G., {Lada} C.~J., 1977, \apj, 214, 725

\bibitem[{{Fabian}(1999)}]{fabian1999}
{Fabian} A.~C., 1999, \mnras, 308, L39

\bibitem[{{Fabian} {et~al.}(2000){Fabian}, {Smail}, {Iwasawa}, {Allen},
  {Blain}, {Crawford}, {Ettori}, {Ivison}, {Johnstone}, {Kneib}, \&
  {Wilman}}]{fabianxray2000}
{Fabian} A.~C. et al., 2000, \mnras, 315, L8

\bibitem[{{Fan} {et~al.}(2001){Fan}, {Narayanan}, {Lupton}, {Strauss}, {Knapp},
  {Becker}, {White}, {Pentericci}, {Leggett}, {Haiman}, Gunn, {Ivezic}, \&
  {Schneider}}]{fan3}
{Fan} X. et al., 2001, AJ, 122, 2833
  astro-ph/0108063

\bibitem[{{Fan} {et~al.}(2000){Fan}, {Strauss}, {Schneider}, {Gunn}, {Lupton},
  {Anderson}, {Voges}, {Margon}, {Annis}, {Bahcall}, {Brinkmann}, {Brunner},
  {Carr}, {Csabai}, {Doi}, {Frieman}, {Fukugita}, {Hennessy}, {Hindsley},
  {Ivezi{\' c}}, {Knapp}, {Lamb}, {McKay}, {Munn}, {Newberg}, {Pauls}, {Pier},
  {Rechenmacher}, {Richards}, {Rockosi}, {Stoughton}, {Szalay}, {Thakar},
  {Tucker}, {Waddell}, \& {York}}]{fan1}
{Fan} X. et al., 2000, \aj, 119, 1

\bibitem[{{Figer} {et~al.}(1998){Figer}, {Najarro}, {Morris}, {McLean},
  {Geballe}, {Ghez}, \& {Langer}}]{figer98}
{Figer} D.~F., {Najarro} F., {Morris} M., {McLean} I.~S., {Geballe} T.~R.,
  {Ghez} A.~M., {Langer} N., 1998, \apj, 506, 384

\bibitem[{{Franceschini} {et~al.}(2000){Franceschini}, {Bassani}, {Cappi},
  {Granato}, {Malaguti}, {Palazzi}, \& {Persic}}]{fran2000}
{Franceschini} A., {Bassani} L., {Cappi} M., {Granato} G.~L., {Malaguti} G.,
  {Palazzi} E., {Persic} M., 2000, \aap, 353, 910

\bibitem[{{Frayer} \& {Brown}(1997)}]{fb97}
{Frayer} D.~T., {Brown} R.~L., 1997, \apjs, 113, 221

\bibitem[{{Friaca} \& {Terlevich}(1998)}]{friacasfr}
{Friaca} A. C.~S., {Terlevich} R.~J., 1998, \mnras, 298, 399

\bibitem[{{Fryer} {et~al.}(2001){Fryer}, {Woosley}, \& {Herger}}]{fryerinpress}
{Fryer} C.~L., {Woosley} S., {Herger} A., 2001, \apj, 550, 372 

\bibitem[{{Granato} {et~al.}(2001){Granato}, {Silva}, {Monaco}, {Panuzzo},
  {Salucci}, {De Zotti}, \& {Danese}}]{granato2001}
{Granato} G.~L., {Silva} L., {Monaco} P., {Panuzzo} P., {Salucci} P., {De
  Zotti} G., {Danese} L., 2001, \mnras, 324, 757

\bibitem[{{Herbig}(1962)}]{herbig62}
{Herbig} G.~H., 1962, \apj, 135, 965

\bibitem[{{Hillebrandt}(1982)}]{hille82}
{Hillebrandt} W., 1982, \aap, 110, L3

\bibitem[{{Hughes} {et~al.}(1998){Hughes}, {Serjeant}, {Dunlop},
  {Rowan-Robinson}, {Blain}, {Mann}, {Ivison}, {Peacock}, {Efstathiou}, {Gear},
  {Oliver}, {Lawrence}, {Longair}, {Goldschmidt}, \& {Jenness}}]{hdfnature}
{Hughes} D.~H. et al., 1998, \nat,
  394, 241

\bibitem[{{Jimenez} {et~al.}(2002){Jimenez}, {Dunlop}, {Padoan}, {Peacock},
  {MacDonald}, \& {J{\o}rgensen}}]{raulssp2002}
{Jimenez} R., {Dunlop} J., {Padoan} P., {Peacock} J., {MacDonald} J.,
  {J{\o}rgensen} U.~G., 2002, \mnras, submitted

\bibitem[{{Jimenez} {et~al.}(1999){Jimenez}, {Friaca}, {Dunlop}, {Terlevich},
  {Peacock}, \& {Nolan}}]{prematuredismissal}
{Jimenez} R., {Friaca} A.~C.~S., {Dunlop} J.~S., {Terlevich} R.~J., {Peacock}
  J.~A., {Nolan} L.~A., 1999, \mnras, 305, L16

\bibitem[{{Jimenez} \& {MacDonald}(1996)}]{rauljmstar2}
{Jimenez} R., {MacDonald} J., 1996, \mnras, 283, 721

\bibitem[{{Jimenez} {et~al.}(2000){Jimenez}, {Padoan}, {Dunlop}, {Bowen},
  {Juvela}, \& {Matteucci}}]{rauldustmodel}
{Jimenez} R., {Padoan} P., {Dunlop} J.~S., {Bowen} D.~V., {Juvela} M.,
  {Matteucci} F., 2000, \apj, 532, 152

\bibitem[{{Jimenez} {et~al.}(1998){Jimenez}, {Padoan}, {Matteucci}, \&
  {Heavens}}]{raulssp1998}
{Jimenez} R., {Padoan} P., {Matteucci} F., {Heavens} A.~F., 1998, \mnras, 299,
  123

\bibitem[{{Jimenez} {et~al.}(1996){Jimenez}, {Thejll}, {J{\o}rgensen},
  {MacDonald}, \& {Pagel}}]{rauljmstar1}
{Jimenez} R., {Thejll} P., {J{\o}rgensen} U.~G., {MacDonald} J., {Pagel} B.,
  1996, \mnras, 282, 926

\bibitem[{Kochanek {et~al.}(2001)Kochanek, Pahre, Falco, Huchra, Mader,
  Jarrett, Chester, Cutri, \& Schneider}]{kochanek2001}
Kochanek C.~S. et al., 2001, ApJ, 560, 566

\bibitem[{{Kormendy} \& {Gebhardt}(2001)}]{kormendy2001}
{Kormendy} J., {Gebhardt} K., 2001, in The 20th Texas Symposium on Relativistic
  Astrophysics, {Martel} H., {Wheeler} J.~C., eds., AIP

\bibitem[{Kukula {et~al.}(2001)Kukula, Dunlop, McLure, Miller, Percival, Baum,
  \& O'Dea}]{kukula2001}
Kukula M.~J., Dunlop J.~S., McLure R.~J., Miller L., Percival W.~J., Baum
  S.~A., O'Dea C.~P., 2001, MNRAS, 326, 1533

\bibitem[{{Lada} {et~al.}(1991){Lada}, {Evans}, {Depoy}, \&
  {Gatley}}]{ladaetal91}
{Lada} E.~A., {Evans} N.~J., {Depoy} D.~L., {Gatley} I., 1991, \apj, 371, 171

\bibitem[{{Laor}(1998)}]{laor98}
{Laor} A., 1998, \apjl, 505, L83

\bibitem[{{Madau} \& {Rees}(2001)}]{madaurees2001}
{Madau} P., {Rees} M.~J., 2001, \apjl, 551, L27

\bibitem[{{McLure} \& {Dunlop}(2002)}]{rossjim2002}
{McLure} R.~J., {Dunlop} J.~S., 2002, MNRAS, 331, 795

\bibitem[{{Merritt} \& {Ferrarese}(2001)}]{mf2001}
{Merritt} D., {Ferrarese} L., 2001, \mnras, 320, L30

\bibitem[{{Miyaji} {et~al.}(2000){Miyaji}, {Hasinger}, \& {Schmidt}}]{mhs1999}
{Miyaji} T., {Hasinger} G.~., {Schmidt} M., 2000, \aap, 353, 25

\bibitem[{{Nomoto} {et~al.}(1984){Nomoto}, {Thielemann}, \&
  {Wheeler}}]{nomoto84}
{Nomoto} K., {Thielemann} F.-K., {Wheeler} J.~C., 1984, \apjl, 279, L23

\bibitem[{{Pentericci} {et~al.}(2001){Pentericci}, {McCarthy}, {R{\"
  o}ttgering}, {Miley}, {van Breugel}, \& {Fosbury}}]{pent2001}
{Pentericci} L., {McCarthy} P.~J., {R{\" o}ttgering} H.~J.~A., {Miley} G.~K.,
  {van Breugel} W.~J.~M., {Fosbury} R., 2001, \apjs, 135, 63

\bibitem[{{Renzini} \& {Voli}(1981)}]{renzini81}
{Renzini} A., {Voli} M., 1981, \aap, 94, 175

\bibitem[{{Scott} {et~al.}(2002){Scott}, {Fox}, {Dunlop}, {Serjeant},
  {Peacock}, {Ivison}, {Oliver}, {Mann}, {Lawrence}, {Efstathiou},
  {Rowan-Robinson}, {Hughes}, {Archibald}, {Blain}, \& {Longair}}]{susiesurvey}
{Scott} S. et al., 2002, MNRAS, 331, 817

\bibitem[{{Silk} \& {Rees}(1998)}]{silkrees1998}
{Silk} J., {Rees} M.~J., 1998, \aap, 331, L1

\bibitem[{{Smail} {et~al.}(2000){Smail}, {Ivison}, {Blain}, {Kneib}, \&
  {Owen}}]{sibk2000}
{Smail} I., {Ivison} R., {Blain} A., {Kneib} J.-P., {Owen} F., 2000, in ASP
  Conf. Ser. 195: Imaging the Universe in Three Dimensions, van Breugel W.,
  Bland-Hawthorn J., eds., p. 248

\bibitem[{{Smail} {et~al.}(1997){Smail}, {Ivison}, \&
  {Blain}}]{smailsurvey1997}
{Smail} I., {Ivison} R.~J., {Blain} A.~W., 1997, \apjl, 490, L5

\bibitem[{{Small} \& {Blandford}(1992)}]{smallblandford92}
{Small} T.~A., {Blandford} R.~D., 1992, \mnras, 259, 725

\bibitem[{{Somerville} \& {Primack}(1999)}]{somervilleprimack1999}
{Somerville} R.~S., {Primack} J.~R., 1999, \mnras, 310, 1087

\bibitem[{{van Breugel} {et~al.}(1998){van Breugel}, {Stanford}, {Spinrad},
  {Stern}, \& {Graham}}]{vbs98}
{van Breugel} W. J.~M., {Stanford} S.~A., {Spinrad} H., {Stern} D., {Graham}
  J.~R., 1998, \apj, 502, 614

\bibitem[{{van Breugel} {et~al.}(1999){van Breugel}, {De Breuck}, {Stanford},
  {Stern}, {Rottgering}, \& {Miley}}]{vanB1999}
{van Breugel} W.~J.~M., {De Breuck} C., {Stanford} S.~A., {Stern} D.,
  {Rottgering} H., {Miley} G.~K., 1999, \apjl, 518, L61

\bibitem[{{Woosley} \& {Weaver}(1995)}]{ww95}
{Woosley} S.~E., {Weaver} T.~A., 1995, \apjs, 101, 181

\bibitem[{{Xu} \& {de Zotti}(1989)}]{xuanddezotti}
{Xu} C., {de Zotti} G., 1989, \aap, 225, 12

\bibitem[{{Zepf}(1997)}]{zepf97}
{Zepf} S.~E., 1997, \nat, 390, 377

\bibitem[{{Zepf} \& {Silk}(1996)}]{zepfandsilk}
{Zepf} S.~E., {Silk} J., 1996, \apj, 466, 114

\end{thebibliography}

\label{lastpage}
\end{document}